\documentclass[a4paper,12pt]{article}

\usepackage[latin1]{inputenc}
\usepackage[american]{babel}

\usepackage{amsfonts}
\usepackage{amsbsy}
\usepackage{amsmath}
\usepackage{epsfig}
\usepackage{hyperref}

 \usepackage{a4wide}
 \usepackage{graphics}
 \usepackage{subfigure}
 \usepackage{fancyhdr}
\newcommand{\der}{\partial}

\newcommand{\ep}{\epsilon}
\newcommand{\half}{\frac{1}{2}}
\newcommand{\bee}{\begin{equation}}
\newcommand{\be}{\begin{equation}}
\newcommand{\ee}{\end{equation}}
\newcommand{\bea}{\begin{eqnarray}}
\newcommand{\eea}{\end{eqnarray}}
\newcommand{\ba}{\begin{eqnarray}}
\newcommand{\ea}{\end{eqnarray}}

\newcommand{\Tr}{\mathrm{Tr}}

\newcommand{\del}{\partial}
\newcommand{\vep}{\varepsilon}
\def\non{\nonumber}
\newcommand{\bPsi}{{\overline \Psi}}
\newcommand{\bep}{{\overline\epsilon}}

\newcommand{\f}[4]{f^{#1 #2 #3}_{\phantom{#1 #2 #3} #4}}

\begin{document}

\begin{titlepage}

\vfill

\begin{flushright}
UB-ECM-PF-08-09\end{flushright}

\vfill

\begin{center}
   \baselineskip=16pt
  {\Large\bf Bagger-Lambert Theory for  General Lie Algebras}
   \vskip 2cm
   Jaume Gomis$^{a}$, Giuseppe Milanesi$^{b}$ and Jorge G.
    Russo$^{b,c}$   
\\
 \vskip .6cm
      \begin{small}
      $^a$\textit{Perimeter Institute for Theoretical Physics\\
      Waterloo, Ontario N2L 2Y5, Canada}
         \end{small}\\*[.5cm]
      \begin{small}
      $^b$\textit{Departament ECM and Institut 
      de Ciencies del Cosmos,\\
Facultat de F\'\i sica, Universitat de Barcelona,\\
Diagonal 647, 08028 Barcelona, Spain}
        \end{small}\\*[.5cm]
      \begin{small}
     $^c$\textit{Instituci\' o Catalana de Recerca i Estudis Avan\c{c}ats (ICREA)}
        \end{small}
   \end{center}
\vfill

\begin{center}
\textbf{Abstract}
\end{center}

\begin{quote}
We construct the totally antisymmetric structure constants $f^{ABCD}$  
of a 3-algebra with a Lorentzian bi-invariant metric
starting from an arbitrary semi-simple Lie algebra.
The structure constants $f^{ABCD}$ can be used to write down a maximally 
superconformal 3d theory that incorporates 
the expected degrees of freedom of multiple M2 branes,
including the ``center-of-mass" mode described by free scalar and fermion fields. 
The gauge field sector reduces to a three dimensional $BF$  term, which underlies 
the gauge symmetry of the theory.
We comment on the issue of unitarity of the quantum theory, which is problematic, despite the fact that the 
specific form of the interactions prevent the ghost fields from running in the internal lines of any Feynman diagram.
Giving an expectation value to one of the scalar fields leads to the maximally supersymmetric 3d  Yang-Mills 
Lagrangian with the addition of two  $U(1)$ multiplets, one of them ghost-like, 
which is decoupled at large $g_{\rm YM}$.
\end{quote}

\vfill

\end{titlepage}
\setcounter{equation}{0}

\section{Introduction}

Finding the three-dimensional superconformal field theory that describes the low energy dynamics of multiple coincident M2 branes
may lead to profound new insights in our understanding of M-theory.
In \cite{Bagger:2007jr} a  maximally supersymmetric three dimensional conformal field theory (henceforth called the BL theory) was proposed as a candidate description of the low energy world volume theory of multiple coincident M2-branes,  incorporating some insights from earlier works \cite{Schwarz:2004yj,Basu:2004ed,Bagger:2006sk}. Some elements of the theory were already present in the important work of Gustavsson
 \cite{Gustavsson:2007vu}.

The BL theory is based on a generalization of  Lie algebras dubbed $3$-algebras\footnote{Known in the mathematical literature as  $3$-Lie algebras \cite{filippov}.}(studied independently by Gustavsson 
in \cite{Gustavsson:2007vu}). A 3-algebra $\mathcal A$ is an $N$  dimensional vector space endowed with a trilinear skew-symmetric product
\begin{equation}
 	[A,B,C]
	\label{threeproduct}
\end{equation}
which satisfies the so called fundamental identity
\begin{equation}\label{absfund}
 	[A,B,[C,D,E]] = [[A,B,C],D,E]+[C,[A,B,D],E]+[C,D,[A,B,E]]\,,
\end{equation}
which extends the familiar Jacobi identity of Lie algebras to $3$-algebras.
If we let $\{T^A\}_{1\leq A\leq N}$ be a basis of $\mathcal A$, the 
$3$-algebra   is specified by the structure constants
$\f ABCD$ of $\mathcal A$:
\begin{equation}
 	[T^A,T^B,T^C] = \f ABCD T^D\,.
\end{equation}
The fundamental identity \eqref{absfund} is expressed as:
\begin{equation}\label{indfund}
 	\f ABGH \f CDEG  = \f ABCG \f GDEH + \f ABDG \f CGEH + \f ABEG \f CDGH\,.
\end{equation}
Classifying 3-algebras $\mathcal A$ requires classifying the solutions to the fundamental identity (\ref{indfund}) for the structure constants $\f ABCD$.

In order to derive from a Lagrangian description the equations of motion of the BL theory -- which were obtained by  demanding closure of the supersymmetry algebra -- a bi-invariant non-degenerate metric $h^{AB}$ on the $3$-algebra $\mathcal A$ is needed. Bi-invariance requires the metric to satisfy:
\begin{equation}\label{absinv}
\f ABCE h^{ED}+\f BCDE h^{AE}=0\,.
\end{equation}
This implies that the  tensor $f^{ABCD}\equiv \f ABCE h^{ED}$ is totally antisymmetric. The metric  $h^{AB}$ arises by postulating a non-degenerate, bilinear  scalar product $\Tr (\,,\,)$ on the algebra
$\mathcal A$:
\begin{equation}
	h^{AB} = \Tr\left(T^A,T^B\right)\,.
\end{equation}

The Lagrangian of the BL theory is  completely specified once a collection of structure constants 
$\f ABCD$  and a bi-invariant  metric $h^{AB}$  solving the constraints  (\ref{indfund}), (\ref{absinv}) is given. The BL theory encodes the interactions of a three dimensional ${\cal N}=8$ multiplet, consisting of eight scalar fields $X^{(I)}$ and their fermionic superpartners $\Psi$,  and   a non-propagating gauge field $A_\mu^{\phantom \mu A}$$_B$. Matter fields in this theory take values in $\mathcal A$, so that $X^{(I)} = X^{(I)}_A T^A, \Psi = \Psi_A T^A$.
The BL Lagrangian is given by \cite{Bagger:2007jr}
\ba
{\cal L} &=& -\half D_\mu X^{A(I)}D^\mu X_A^{(I)} +
\frac{i}{2}{\bPsi}^A\Gamma^\mu D_\mu \Psi_A
+\frac{i}{4} f_{ABCD} \bPsi^B \Gamma^{IJ}X^{C(I)}X^{D(J)}\Psi^A\non 
\\
 &-&\frac{1}{12} \left(f_{ABCD}X^{A(I)}X^{B(J)} X^{C(K)}\right)
\left(f_{EFG}^{\phantom{EFG}D}X^{E(I)}X^{F(J)}X^{G(K)} \right)\non \\
&+&\half\,\vep^{\mu\nu\lambda}\left(
f_{ABCD}A_\mu^{~AB}\del_\nu A_{\lambda}^{~CD}
+ \frac23 f_{AEF}^{~~~~~G}\,f_{BCDG}\,
A_{\mu}^{~AB}A_{\nu}^{~CD}A_{\lambda}^{~EF}\right),
\label{lba}
\ea
where:
\begin{equation}
D_\mu \Phi^{A(I)} = \del_\mu \Phi^{A(I)} + f^{A}_{~~BCD}A_\mu^{CD}\Phi^{B(I)}\;.
\end{equation}
The theory is invariant under the gauge transformations
\ba
\delta X^{A(I)} &=& -f^A_{~~BCD}\Lambda^{BC}X^{D(I)} \non\\
\delta \Psi^A &=& -f^A_{~~BCD}\Lambda^{BC}\Psi^D\non\\
\delta (f_{AB}^{~~~\,CD}A_\mu^{AB}) &=& f_{AB}^{~~~\,CD}D_\mu \Lambda^{AB} 
\ea
and under the  following supersymmetry transformations
\ba
\label{sus}
\delta X^{A(I)} &=& i\,\bep\,\Gamma^I\Psi^A \non\\
\delta \Psi^A & =& D_\mu X^{A(I)}\Gamma^\mu \Gamma^I \ep +\frac16
f^A_{~~BCD}X^{B(I)}X^{C(J)}X^{D(K)} \Gamma^{IJK}\ep \non\\
\delta (f_{AB}^{~~~\,CD}A_\mu^{AB}) &=& i f_{AB}^{~~~\,CD}
X^{A(I)}\,\bep\, \Gamma_\mu\Gamma_I
\Psi^B,
\ea
where  $\Psi$  and $\ep$ are eleven dimensional Majorana spinors satisfying the projection condition
$\Gamma_{012}\ep = \ep$ and $\Gamma_{012}\Psi^A = -\Psi^A$ respectively.

\medskip

{}The only non-trivial example of  a $3$-algebra with a positive definite 
$3$-algebra metric  $h^{AB}$  is the four dimensional algebra $\mathcal A_4$, defined by structure constants $\f ABCD={\vep ^{ABC}}_D$, where $\epsilon^{ABCD}$ 
is the 4-dimensional Levi Civita symbol. In \cite{Papadopoulos:2008sk,Gauntlett:2008uf} it has been proven  that the only $3$-algebras with a positive definite 
$3$-algebra metric  $h^{AB}$ are   
${\cal A}_4\oplus \ldots\oplus {\cal A}_4\oplus C_1\oplus\ldots \oplus C_l$, where $C_i$ denote central elements in the algebra\footnote{As previously conjectured in e.g. \cite{Gustavsson:2008dy,Ho:2008bn}.}.   
New constructions are possible if one does not require the existence of a Lagrangian but 
only of the equations of motion \cite{Gran:2008vi}, which can be 
written without the need of a metric $h^{AB}$ in the algebra.

In this paper we find a novel construction of $3$-algebras ${\cal A}_{\mathfrak  g}$  based of an arbitrary semi-simple Lie algebra ${\mathfrak  g}$,   giving rise to an infinite class of novel realizations of the BL theory. These new $3$-algebras are found by relaxing the condition that the  $3$-algebra metric $h^{AB}$ is positive definite\footnote{Earlier studies of 3-algebras for Lorentzian metrics can be found in \cite{FigueroaO'Farrill:2002xg}.}. 
In our construction the  $3$-algebra metric is taken to be $h^{AB}={\rm diag}(-1,1,...,1)$, and it has a single  timelike direction.

In most physical theories, a positive-definite metric is required in order to ensure that the theory has 
positive-definite kinetic terms
  and to 
 prevent  violations of unitarity due to propagating ghost-like degrees of freedom.
Nevertheless,  there are examples of theories that are unitary  despite the presence of ghost fields, 
like Chern-Simons theory based on non-compact Lie algebras \cite{witten1,witten2}.
The peculiar form of the interactions  make our model  resemble, in some aspects, the Nappi-Witten model \cite{nappi}, describing a WZW model for  a non semi-simple algebra, and analogous constructions for Chern-Simons and Yang-Mills theories in \cite{tseytlin} based on non semi-simple gauge groups.

The BL theory was considered recently in several papers. Full superconformal invariance was proven in \cite{Bandres:2008vf}. 
In \cite{Mukhi:2008ux} a specific way to connect the BL theory to the 
D2-brane theory by giving a vacuum expectation value to a scalar field was proposed.
Different discussions of the vacuum moduli space  appeared in 
\cite{Bagger:2007vi,VanRaamsdonk:2008ft,Lambert:2008et,Distler:2008mk}. 
The proposal seems to be that the
BL theory with algebra ${\cal A}_4 $ describes two M2-branes propagating in a non trivial orbifold of flat space.
A maximally supersymmetric deformation of the theory by a mass parameter was found in \cite{Gomis:2008cv,Hosomichi:2008qk}.
In \cite{Bergshoeff:2008cz}  it was shown that the BL theory fits in the general construction of maximally supersymmetric gauge theories using the embedding tensor techniques.
Other   interesting recent papers on BL theory have appeared in \cite{Berman:2008be,Morozov:2008cb,Ho:2008nn}.


\section{The model}

We take the  bi-invariant metric on the $3$-algebra ${\cal A}$ to be
\be
h^{AB}=\eta^{AB}\ ,\qquad A,B=0,1,...,n+1,
\ee
where $N=n+2$ is the dimension of ${\cal A}$ and $\eta^{AB}={\rm diag}(-1,1,...,1)$ is the Minkowski metric on the $3$-algebra ${\cal A}$.

We now split the   $3$-algebra indices $A,B,...$ into $A=(0,a,\phi)$ where
$a,b=1,...,n$ and $\phi\equiv n+1$.
Then the following set of totally antisymmetric structure constants
\be
f^{0 abc}=f^{\phi abc}=C^{abc}\ ,\qquad f^{0\phi ab}=f^{abcd}=0\ ,
\ee
solve the fundamental identity (\ref{indfund}), where 
$C^{abc}$ are the structure constants of a compact semi-simple Lie algebra
${\mathfrak  g}$ of dimension $n$. The  structure constants $C^{abc}$ satisfy the usual Jacobi identity.

Therefore, for any given semi-simple Lie algebra ${\mathfrak  g}$,
one can construct an associated $3$-algebra, which we will denote by  ${\cal A}_{\mathfrak  g}$. This means that we can write down an explicit realization of the Bagger-Lambert theory for any  semi-simple Lie algebra ${\mathfrak  g}$. This gives rise to a family of maximally supersymmetric Lagrangians in three dimensions.

It  is convenient to introduce ``light-cone variables", that is  null generators  on the
algebra
${\cal A}_{\mathfrak  g}\;$:
\be
T^{\pm}=\pm T^0+T^\phi.
\ee
In this basis the metric in ${\cal A}_{\mathfrak  g}$ is given by
\be
h^{+-}=2,\qquad h^{\pm\pm}=0,\qquad h^{ab}=\delta^{ab},\qquad h^{a\pm }=0\ ,
\ee
while the structure constants of ${\cal A}_{\mathfrak  g}$ are given by:
 \be
f^{+abc}=2 C^{abc}\ ,\qquad f_{-abc}= C_{abc} \ ,\qquad f^{-abc}=f_{+abc}=0\,.
\label{algebranuestra}
\ee

In order to write the Lagrangian we define $X^{\pm(I)}=\pm X^{0(I)}+X^{\phi (I)}$ and $\Psi^{\pm}=\pm \Psi^{0}+\Psi^{\phi}$. The Lagrangian based on   ${\cal A}_{\mathfrak  g}$ now reads
\ba
{\cal L} &=&  -\half (\del_\mu X^{+(I)} + 4B_{\mu a} X^{a(I)})
\del^\mu X^{-(I)} 
-\half D_\mu X^{a(I)}D^\mu X_a^{(I)} 
\non\\
&+&
\frac{i}{2}{\bPsi}^a\Gamma^\mu D_\mu \Psi_a
+\frac{i}{4} {\bPsi}^+\Gamma^\mu \del_\mu \Psi^-+
\frac{i}{4} {\bPsi}^-\Gamma^\mu (\del_\mu \Psi^+ + 4B_{\mu a}\Psi^a )
\non\\
&+&\frac{i}{2} C_{abc} \bPsi^a \Gamma^{IJ}X^{b(I)}X^{c(J)}\Psi^-
+\frac{i}{2} C_{abc} \bPsi^b \Gamma^{IJ}X^{-(I)}X^{c(J)}\Psi^a
\non \\
&-&\frac{1}{4} \left(C_{abc}X^{a(I)}X^{b(J)} X^{-(K)}\right)
\left(C_{ef}^{\ \ \ c} X^{e(I)}X^{f(J)}X^{-(K)} \right)\non\\
&-&\frac{1}{2} \left(C_{abc}X^{a(I)}X^{b(J)} X^{-(K)}\right)
\left(C_{fe}^{\ \ \ c} X^{e(I)}X^{f(K)}X^{-(J)} \right)
\non \\
&+&2 \vep^{\mu\nu\lambda}
B_\mu^{~a} F_{\nu \lambda }^{~a}\;,
\label{modelo}
\ea
where we have decomposed the gauge fields    as follows 
\be
A^a_\mu \equiv A^{-a}_\mu\ ,\qquad  B^a_\mu\equiv{1\over 2} C^{abc} A_{\mu bc}\;,
\ee
the curvature is given by 
\be
F_{\nu \lambda }^{~a}= \del_\nu A_\lambda^ a - \del_\lambda A_\nu^ a  - 2 C^a_{\ \ bc} A^b_\nu A^c_\lambda 
\ee
and:
\be
D_\mu X^{a(I)} = \del_\mu X^{a(I)} -2 B^{a}_\mu X^{-(I)}
+2 C^{a}_{~~bc}A_\mu^{c}X^{b(I)}\,.
\ee
We note  that the gauge  fields $A^{+-}_\mu $ and $A^{+b}_\mu $ do not appear in the Lagrangian, gauge transformations and supersymmetry transformations. Therefore,   they are not part of the theory.
Similarly, $A_{\mu bc} $ appears only through the combination $C^{abc} A_{\mu bc}=2B_{\mu}^a$,
so $A_\mu^a, \ B_\mu^a$ will be viewed as the fundamental gauge fields in the theory.
The Bagger-Lambert Chern-Simons term reduces, in our case, to a three dimensional $BF$ term.

It should be noted that structure constants defined by introducing an overall  multiplicative parameter $\kappa^2 $, i.e. $f^{+abc}=2 \kappa^2 C^{abc}$, also solve the fundamental identity. 
Importantly, $\kappa^2 $ can be rescaled away from the Lagrangian by rescaling
$X^a\to X^a, X^-\to  X^-/\kappa^2,\ X^+\to  \kappa^2 X^+,\ B^a_{\mu}\to \kappa^2 B^a_{\mu}, A_\mu^a\to A_\mu^a/\kappa^2$, and similarly for the fermion fields. \footnote{The fact that $\kappa^2 $ can be rescaled away  was first noticed in \cite{verlinde,matsuo}.}

The Lagrangian (\ref{modelo}) is invariant under the following gauge transformations  
\ba
 \delta B_\mu^ c  &=&\del_\mu \tilde{\Lambda} ^{c}-2C^{c}_{\ \ ab} B^{a}_\mu \Lambda^{b}-2C^{c}_{\ \ da} A_\mu^d\tilde{\Lambda} ^{a}
 \non\\
\delta A_{\mu }^a &=& \del_\mu \Lambda^{a}+2 C^a_{\ \ bc}A^c_\mu \Lambda^{b}
\non\\
\delta X^{a(I)} &=& 2 \tilde{\Lambda}^{a}X^{-(I)} + 2 C^a_{\ \  bc} \Lambda^{b}X^{c(I)}
\non\\
\delta X^{+(I)} &=& -4  \tilde{\Lambda}_cX^{c(I)} \non\\
 \delta X^{-(I)} &=&0
\non\\
\delta \Psi^{a} &=&2 \tilde{\Lambda}^{a}\Psi^{-} + 2 C^a_{\ \  bc} \Lambda^{b}\Psi^{c}
\non\\
\delta \Psi^{+} &=& -4  \tilde{\Lambda}_c\Psi^{c}\non\\
 \delta \Psi^{-} &=& 0 
\ea
where $\Lambda^ a \equiv \Lambda^{-a} $ and $ \tilde \Lambda^{a}\equiv \half C^a_{\ \ bc}\Lambda^{bc}$.
The supersymmetry transformations are given by
\ba
\label{susyB}
\delta X^{A(I)} &=& i\,\bep\,\Gamma^I\Psi^A\ ,\qquad A=\{-,+,a\} 
\non\\
\delta \Psi^- & =& \del_\mu X^{-(I)}\Gamma^\mu \Gamma^I \ep 
\non\\
\delta \Psi^+ & =& (\del_\mu X^{+(I)}+4 B_{\mu a} X^{a(I)}) \Gamma^\mu \Gamma^I \ep +{1\over3}
C^{bcd}X^{b(I)}X^{c(J)}X^{d(K)} \Gamma^{IJK}\ep 
\non\\
\delta \Psi^a & =& D_\mu X^{a(I)}\Gamma^\mu \Gamma^I \ep -\half
C^a_{~~bc}X^{b(I)}X^{c(J)}X^{-(K)} \Gamma^{IJK}\ep 
\non\\
\delta B_\mu^{c} &=& {i\over2}\ C_{ab}^{~~\,c}
X^{a(I)}\,\bep\, \Gamma_\mu\Gamma_I
\Psi^b
\non\\ 
\delta A_\mu^{a} &=&  {i\over 2}
X^{-(I)}\,\bep\, \Gamma_\mu\Gamma_I
\Psi^a -{ i\over 2} 
X^{a(I)}\,\bep\, \Gamma_\mu\Gamma_I
\Psi^-\;.
\label{susynueva}
\ea
A remarkable feature of the  Lagrangian (\ref{modelo}) is that the classical equations
of motion for $X^{+(I)}, \Psi^{+}$ imply that:
\be
\del_\mu \del^ \mu X^ {-(I)} = 0\ ,\qquad \Gamma^ \mu \del_\mu \Psi^- = 0\;.
\label{asa}
\ee
Therefore,  $X^ {-(I)} $ and  $\Psi^{-} $ propagate as  free fields (even though they participate in interactions). 

\medskip

The Lagrangian can also be understood as an ordinary gauge theory (with an invariant metric) for an  ``extended" Lie algebra ${\cal G}$. The Lie algebra ${\cal G}$ is generated by $S^{AB}$, whose matrix elements are given by $\left( S^{AB}\right)^C{}_D = \f ABCD$  \cite{Gustavsson:2007vu}
(the fundamental identity \eqref{indfund} indeed implies that the matrices $(S^ {AB})^C_{\ D}$ generate a Lie algebra ${\cal  G}$).
The structure is as follows (see appendix for more details).
A generic element of $\mathcal G$ is determined by an antisymmetric matrix $\Omega_{AB} = -\Omega_{BA}$ and the action of 
$L(\Omega_{AB})\in\mathcal G$ on $\mathcal A$ is given by:
\begin{equation}
 	 L(\Omega_{AB}) \cdot T^C = \Omega_{AB} [T^A,T^B,T^C] = \Omega_{AB} \f ABCD T^D\,.
\end{equation}
For our $3$-algebra ${\cal A}_{\mathfrak  g}$ (\ref{algebranuestra}), the explicit form of the generators 
of $\mathcal G$
is given by:
\be
(J^a)^B_{\ C} =- \half (S^{+a})^B_{\ C}\ ,\qquad  (P^a)^B_{\ C} =  2 \delta^{a}_C\delta^B_+ - \delta^{aB} \delta^-_C
= {1\over c_2} {C^a}_{de} f^{deB}_{\ \ \ \ C}\, ,
\ee
where we have used $C^a_{\ \ cd} C^{bcd }= c_2 \delta^{ab}\ $ and $c_2$ is the quadratic Casimir in the adjoint of ${\mathfrak  g}$.

Hence the algebra ${\cal G}$ has dimension $ {\rm dim}\ {\cal G}= 2n$.
The generators of ${\cal G}$  obey the following commutation relations:
\be
[P^a,P^b]=0\ ,\qquad [J^a, J^b]={C^{ab}}_c J^c\ ,\qquad [P^a,J^b] = 
{C^{ab}}_c P^c\,.
\label{aaoo}
\ee
The algebra (\ref{aaoo}) is recognized as the symmetry algebra of  three dimensional $BF$
theories \cite{witten1} (a review on $BF$ theory can be found in \cite{blau}).  
${\cal G}$ has the structure of a semi-direct sum of $n$ abelian generators with a semi-simple Lie algebra  ${\mathfrak  g}$. More precisely, it is the semi-direct sum of the translation algebra with  ${\mathfrak  g}$.
The $B_\mu^a$ and $A_\mu^a$ gauge fields are associated with the generators $P^a$ and $J^a$ respectively. For the case  ${\mathfrak  g}=su(2)$, the extended Lie algebra ${\cal G}$   is
the Lie algebra $iso(3)$, where the generators $P^a$ are associated with translations while the generators
$J^a$ are associated with $so(3)=su(2)$ rotations\footnote{
One could choose ${\mathfrak  g}=so(2,1)$ to obtain a theory (\ref{modelo}) containing the Lagrangian of three dimensional gravity \cite{witten1} coupled to matter in a way that  
$iso(2,1)$ gauge invariance is maintained, even though 
it is not invariant under diffeomorphisms.}. The generators in this representation
are explicitly given in the appendix.

\medskip

In the quantum theory, the path integral over  $X^{+(I)}, \Psi^{+}$ completely freezes 
 the modes of 
 $X^{-(I)}, \Psi^{-}$ to their free field values.
This is very similar to what happens for pp wave string models, or 
for WZW models based on non semi-simple Lie algebras   \cite{nappi}. Theories with similar features based on non semi-simple Lie algebras 
have been constructed for Chern-Simons and  Yang-Mills theories \cite{tseytlin}.
These theories have the remarkable property of being one-loop exact. The key mechanism
that takes place is the following.
 Since  
one of the light-cone variables, say $X^+$, does not appear in the interaction vertices and there is no
$X^-X^-$ propagator, there is no  Feynman diagram that one can draw beyond one loop.
This has been used in \cite{nappi} to show that a certain plane wave model is an exact conformal field theory 
and in \cite{tseytlin} to show the remarkable fact that in these types of Yang-Mills theories the on-shell
scattering amplitudes are finite.

An important difference with the present theory is that, although  there are no internal lines in Feynman diagrams involving $X^{\pm (I)}$ and $\Psi^{\pm }$, there are extra fields that can
run in the loop diagrams. 
Another difference arises in the gauge field sector. Because of the peculiar form of the Bagger-Lambert Chern-Simons term in (\ref{lba}) -- where the kinetic term is contracted with  the structure constants -- the field $A^ {+a}_\mu$ does not appear in the Lagrangian
(recall that $f_{+abc}=0$). As a result, since there is no analogue of the equation of motion for $A^ {+a}_\mu$, there is no condition that freezes out the mode  $A^ {-a}_\mu$
as in  (\ref{asa}). 
Nevertheless in the pure $BF$ sector the theory is unitary.

Therefore the quantum interactions in the present theory are non-trivial and, as in ${\cal N}=4$ SYM, we expect contributions from all loops to a generic observable.
It seems possible that  quantum interactions can  be simplified for a suitable gauge fixing, due to
the special nature of $BF$ theories.

\medskip


\section{Connecting to $D2$-branes}

In this section we show how the theory, if interpreted as a theory of coinciding membranes, can be connected to the low energy description of multiple D2 branes.
We follow a similar strategy as in \cite{Mukhi:2008ux}, by giving an expectation value
to one of the scalar fields.
In the present case we propose that
\be
\langle X^{-(8)}\rangle = v\ ,
\label{vvk}
\ee
and zero for all other fields. In general, the fundamental identity implies that the structure constants 
$f^{\alpha AB}_{\ \ \ \ C}$, where $\alpha$ labels an arbitrary $3$-algebra generator, 
satisfy the usual Jacobi identity. Therefore $f^{\alpha AB}_{\ \ \ \ C}$ are
the structure constants of a  conventional Lie algebra.
In the present case of our $3$-algebra  ${\cal A}_{\mathfrak  g}$ (\ref{algebranuestra}) and taking $\alpha=+$, 
the ``reduced" algebra is ${\mathfrak  g} \times u(1)$.

 We now expand the Lagrangian (\ref{modelo})  around the VEV (\ref{vvk}) and identify $g_{\rm YM}=v $.
  As in \cite{Mukhi:2008ux}, we will neglect terms which are suppressed
by powers of $1/g_{\rm YM}$ compared to the leading terms.
For the part involving $B_\mu^a$, we find
\be
{\cal L}_B =  - 2 g_{\rm YM}^2
B_{\mu a} B^{\mu a} +2 g_{\rm YM}B^{\mu a}D'_\mu X_a^{(8)} 
+  2\vep^{\mu\nu\lambda}B_\mu^{~a} F_{\nu \lambda }^{~a}+...
\ee
where $D'_\mu X^{a(I)}=\del_\mu X^{a(I)} -2 C^a_{\ \ bc}A_\mu^b X^{c(I)}$, and the dots represent
terms which give suppressed contributions.
We eliminate $B_\mu^a$ by its equation of motion:
\be
B^a_\mu ={1\over 2g_{\rm YM}^2} \vep_\mu^{\ \ \nu\lambda}F^a_{\nu\lambda} +{1\over 2g_{\rm YM}
}D'_\mu X^{a(8)}\,.
\ee
Inserting this back 
into the Lagrangian, and rescaling $A_\mu^a\to A_\mu^a/2$, we get as leading term in $g^2_{\rm YM}$ the three dimensional SYM Lagrangian
\ba
{\cal L}&=& -{1\over 4 g_{\rm YM}^2} F^a_{\mu\nu}  F^{\mu\nu}_a -\half \del_\mu X^{+(I)} \del^\mu X^{-(I)} 
-\half D_\mu X^{a(i)}D^\mu X_a^{(i)} 
\non\\
&+&
\frac{i}{2}{\bPsi}^a\Gamma^\mu D_\mu \Psi_a
+\frac{i}{2} {\bPsi}^+\Gamma^\mu \del_\mu \Psi^-+
\frac{i}{2} {\bPsi}^-\Gamma^\mu \del_\mu \Psi^+ 
\non\\
&+& g_{\rm YM} \frac{i}{2} C_{abc} \bPsi^b \Gamma^{8j} X^{c(j)}\Psi^a
-\frac{g_{\rm YM}^2}{4} \left(C_{abc}X^{a(i)}X^{b(j)} \right)
\left(C_{ef}^{\ \ \ c} X^{e(i)}X^{f(j)} \right)\;,
\ea
where $i,j=1,...,7$. We also note that the supersymmetry transformations in (\ref{susynueva}) reduce to those of three dimensional ${\cal N}=8$ SYM to leading order in $g_{\rm YM}$ (with $\Gamma^8 $ playing the role of $\Gamma^{10}$).

We can dualise the scalar $X^{\phi(8)}$  by   abelian duality to produce a
$U(1)$ gauge field, and the $U(1)$ supermultiplet is completed by $X^{\phi(i)}$, $\Psi^\phi$.
Taking ${\mathfrak  g} =su(N)$,
the resulting theory is the maximally supersymmetric $SU(N)\times U(1)$ Yang-Mills theory
plus an additional $U(1)$ supermultiplet of free ghost fields,
\be
{\cal L}_{\rm ghost}= {1\over 4} F_{\mu\nu}^2 + \half \del_\mu X^{0(i)} \del^\mu X^{0(i)}- \frac{i}{2} 
{\bPsi}^0\Gamma^\mu \del_\mu \Psi^0 
\ee
where we have dualised $X^{0(8)}$ into an abelian vector field $A_\mu$.
In this limit the ghost Lagrangian is completely decoupled from the $SU(N)\times U(1)$ Yang-Mills theory and it does not affect its unitarity.

A similar theory with a decoupled U(1) ghost has been considered by Tseytlin \cite{tseytlin}. The starting point is 
 $SU(2)\times U(1)$ YM theory with a decoupled-ghost $U(1)$ field. By a contraction of  $SU(2)\times U(1)$
 one ends up with YM theory based on the 4-dimensional  non semi-simple Lie algebra $E_2^c$. It would be interesting
to see if similar limits can be taken at the level of the 3-algebra studied here.

\section{Concluding Remarks}

In general,  the presence of ghost-like particles  renders a theory potentially
non-unitary. 
There are some special cases like Chern-Simons theory based on non-compact semi-simple algebras
where one can show that the theory is nevertheless unitary \cite{witten2}.
Although the present theory also has Chern-Simons gauge fields, there
are some important differences,
in particular, there are extra propagating ghost-like degrees of freedom $X^{0(I)},\ \Psi^{0}$. 
Clearly, in order to settle the unitarity issue, the 
theory requires a separate and more detailed study. 

An interesting feature is that the $X^{+(I)},\ \Psi^+ $ fields can be integrated out exactly,
 freezing out the modes $X^{-(I)},\ \Psi^-$ to their free theory values.
This property ensures that there are modes which may potentially describe the center-of-mass 
translational mode
of multiple M2 branes. In addition,
the fact that interactions only involve $X^{-(I)},\Psi^-$, and not $X^{+(I)},\Psi^+$, implies that no ghost-like $X^{0(I)} , \Psi^0$ field ever appears in internal lines of Feynman diagrams.

It would  also be interesting to see if the present theory could represent multiple M2 branes, if not in a fundamental sense, at least as an 
effective description (e.g. large $N$, where the ghost contributions of $O(1)$ are negligible compared to $N$).

In conclusion, a family of maximally supersymmetric conformal field theories with a Lagrangian formulation exist, and with arbitrary Lie algebra structure.
Their  relevance for M-theory remains to be seen.

\bigskip

\noindent {\bf Note added:} After this paper appeared,  two other papers 
with closely related results \cite{verlinde,matsuo}  appeared in the arXiv.

\section*{Acknowledgments}

J.G. would like to thank L. Freidel for useful discussions and the University of Barcelona for hospitality. G. M. would like to thank M. Gaberdiel for enlightening discussions. J.R. would like to thank P. Townsend and A. Tseytlin for useful comments and the Perimeter Institute for hospitality during the course of this work.
Research at Perimeter Institute is supported by the Government
of Canada through Industry Canada and by the Province of Ontario through
the Ministry of Research and Innovation. J.G.  also acknowledges further  support by an NSERC Discovery Grant.
J.R. acknowledges support by MCYT FPA 2007-66665, European
EC-RTN network MRTN-CT-2004-005104 and CIRIT GC 2005SGR-00564.

\section{Appendix: Induced Lie algebra structure}

In the  examples we constructed, the algebra $\mathcal G$ is determined by $\mathfrak g$. In particular, we will show that $\mathcal G $ is a semidirect sum of $\mathfrak g$  with $n$ abelian generators.
The set $S^{AB}$ of generators of $\mathcal G$ have the following matrix representation which acts on $\mathcal A$ itself:
\begin{equation}
	\left( S^{AB}\right)^C{}_D = \f ABCD\,.
\end{equation}
In our case the $S^{-A}$ generators vanish. The remaining generators are given by
\begin{equation}
	\left( J^{a}\right)^B {}_C \equiv -
	\half \left( S^{+a}\right)^B {}_C = -C^{aB}{}_C\qquad 
	\left(H^{ab}\right)^C{}_D = 2 C^{ab}{}_D \delta^C_+ - C^{abC}\delta^{-}_D\,,
\end{equation}
with $C^{ab\pm} = C^{a+-}=0$. Since $\mathfrak g$ is semisimple, the $J^{a}$ generators are linearly independent.
One can easily check by direct calculation that the $H^{ab}$ generators are abelian. In principle, there are $\frac 12 n(n-1)$ such generators (we recall that $n$ is the dimension of $\mathfrak g$), but each matrix $H^{ab}$ has non vanishing entries only in the $+$ row   and in the $-$ column   (which are proportional).  As such, at most $n$ of them are linearly independent and, due to the fact that $\mathfrak g$ is semisimple, exactly $n$ of them are linearly independent. We can write a basis of the space spanned by $H^{ab}$ as:
\begin{equation}
	\left( P^a \right)^C{}_D = 2 \delta^{a}_D\delta^C_+ - \delta^{aC} \delta^-_D\,.
\end{equation}
A straightforward calculation gives:
\begin{equation}
	[P^a,P^b]=0\ ,\qquad [J^a, J^b]={C^{ab}}_c J^c\ ,\qquad [P^a,J^b] = 
{C^{ab}}_c P^c\,.
\end{equation}
The generic covariant derivative is given by
\begin{equation}
	D_\mu \phi^A = \der_\mu \phi^A + \f CDAB A_{\mu\, CD}\,\phi^B\,.
\end{equation}
Recalling the definitions
\begin{equation}
	A^a_\mu \equiv A^{-a}_\mu\ ,\qquad  B^a_\mu\equiv{1\over 2} C^{abc} A_{\mu bc}\;,
\end{equation}
we have
\begin{equation}
	D_\mu \phi^A = \der_\mu \phi^A + 2A^a_\mu \left(J_a\right)^A{}_B\phi^B + 2 B_\mu^a \left(P_a\right)^A{}_B\phi^B
\end{equation}
which is the standard covariant derivative, as appeared in section 2.

As an example, we explicitly write down the generators of ${\cal G}$ for the simple case in which $\mathfrak g = su(2)$, so that the dimension of ${\cal A}_{\mathfrak  g}$ is $N=5$:
\begin{equation}
 	J^1 = \begin{pmatrix}
 	       	0 & 0 & 0 & 0 & 0\\
	        0 & 0 & 0 & 0 & 0\\
		0 & 0 & 0 & 0 & 0\\
		0 & 0 & 0 & 0 & -1\\
		0 & 0 & 0 & 1 & 0
 	      \end{pmatrix}
\qquad
 	J^2 = \begin{pmatrix}
 	       	0 & 0 & 0 & 0 & 0\\
	        0 & 0 & 0 & 0 & 0\\
		0 & 0 & 0 & 0 & 1\\
		0 & 0 & 0 & 0 & 0\\
		0 & 0 & -1 & 0 & 0
 	      \end{pmatrix}
\qquad
J^3 = \begin{pmatrix}
 	       	0 & 0 & 0 & 0 & 0\\
	        0 & 0 & 0 & 0 & 0\\
		0 & 0 & 0 & -1 & 0\\
		0 & 0 & 1 & 0 & 0\\
		0 & 0 & 0 & 0 & 0
 	      \end{pmatrix}
\end{equation}
and
\begin{equation}
 	P^1 = \begin{pmatrix}
 	       	0 & 0 & 2 & 0 & 0\\
	        0 & 0 & 0 & 0 & 0\\
		0 & -1 & 0 & 0 & 0\\
		0 & 0 & 0 & 0 & 0\\
		0 & 0 & 0 & 0 & 0
 	      \end{pmatrix}
\qquad
 	P^2 = \begin{pmatrix}
 	       	0 & 0 & 0 & 2 & 0\\
	        0 & 0 & 0 & 0 & 0\\
		0 & 0 & 0 & 0 & 0\\
		0 & -1 & 0 & 0 & 0\\
		0 & 0 & 0 & 0 & 0
 	      \end{pmatrix}
\qquad
P^3 = \begin{pmatrix}
 	       	0 & 0 & 0 & 0 & 2\\
	        0 & 0 & 0 & 0 & 0\\
		0 & 0 & 0 & 0 & 0\\
		0 & 0 & 0 & 0 & 0\\
		0 & -1 & 0 & 0 & 0
 	      \end{pmatrix}
\end{equation}
They assemble to build the algebra of $iso(3)$, where the $P^1,\ P^2,\ P^3 $ generate translations and
the $J^1, J^2,\ J^ 3$ generate rotations.


\begin{thebibliography}{999}

\bibitem{Bagger:2007jr}
  J.~Bagger and N.~Lambert,
  ``Gauge Symmetry and Supersymmetry of Multiple M2-Branes,''
  Phys.\ Rev.\  D {\bf 77} (2008) 065008
  [arXiv:0711.0955 [hep-th]].

\bibitem{Schwarz:2004yj}
  J.~H.~Schwarz,
  ``Superconformal Chern-Simons theories,''
  JHEP {\bf 0411} (2004) 078
  [arXiv:hep-th/0411077].

\bibitem{Basu:2004ed}
  A.~Basu and J.~A.~Harvey,
  ``The M2-M5 brane system and a generalized Nahm's equation,''
  Nucl.\ Phys.\  B {\bf 713} (2005) 136
  [arXiv:hep-th/0412310].

\bibitem{Bagger:2006sk}
  J.~Bagger and N.~Lambert,
  ``Modeling multiple M2's,''
  Phys.\ Rev.\  D {\bf 75} (2007) 045020
  [arXiv:hep-th/0611108].


\bibitem{Gustavsson:2007vu}
  A.~Gustavsson,
  ``Algebraic structures on parallel M2-branes,''
  arXiv:0709.1260 [hep-th].

\bibitem{filippov}
V.T. Filippov, ``n-Lie algebras," Sib. Mat. Zh., 26 No. 6, 126140 (1985).


\bibitem{Papadopoulos:2008sk}
  G.~Papadopoulos,
  ``M2-branes, 3-Lie Algebras and Plucker relations,''
  arXiv:0804.2662 [hep-th].

\bibitem{Gauntlett:2008uf}
  J.~P.~Gauntlett and J.~B.~Gutowski,
  ``Constraining Maximally Supersymmetric Membrane Actions,''
  arXiv:0804.3078 [hep-th].

\bibitem{Gustavsson:2008dy}
  A.~Gustavsson,
  ``Selfdual strings and loop space Nahm equations,''
  arXiv:0802.3456 [hep-th].

\bibitem{Ho:2008bn}
  P.~M.~Ho, R.~C.~Hou and Y.~Matsuo,
  ``Lie 3-Algebra and Multiple M2-branes,''
  arXiv:0804.2110 [hep-th].

\bibitem{FigueroaO'Farrill:2002xg}
  J.~Figueroa-O'Farrill and G.~Papadopoulos,
  ``Pluecker-type relations for orthogonal planes,''
  arXiv:math/0211170.


\bibitem{Gran:2008vi}
  U.~Gran, B.~E.~W.~Nilsson and C.~Petersson,
  ``On relating multiple M2 and D2-branes,''
  arXiv:0804.1784 [hep-th].



\bibitem{witten1}
  E.~Witten,
  ``(2+1)-Dimensional Gravity as an Exactly Soluble System,''
  Nucl.\ Phys.\  B {\bf 311}, 46 (1988).

\bibitem{witten2}
  D.~Bar-Natan and E.~Witten,
  ``Perturbative expansion of Chern-Simons theory with noncompact gauge
  group,''
  Commun.\ Math.\ Phys.\  {\bf 141}, 423 (1991).


\bibitem{nappi}
  C.~R.~Nappi and E.~Witten,
  ``A WZW model based on a nonsemisimple group,''
  Phys.\ Rev.\ Lett.\  {\bf 71}, 3751 (1993)
  [arXiv:hep-th/9310112].

\bibitem{tseytlin}
  A.~A.~Tseytlin,
  ``On gauge theories for nonsemisimple groups,''
  Nucl.\ Phys.\  B {\bf 450}, 231 (1995)
  [arXiv:hep-th/9505129].


\bibitem{Bandres:2008vf}
  M.~A.~Bandres, A.~E.~Lipstein and J.~H.~Schwarz,
  ``N = 8 Superconformal Chern--Simons Theories,''
  arXiv:0803.3242 [hep-th].

\bibitem{Mukhi:2008ux}
  S.~Mukhi and C.~Papageorgakis,
  ``M2 to D2,''
  arXiv:0803.3218 [hep-th].


\bibitem{Bagger:2007vi}
  J.~Bagger and N.~Lambert,
  ``Comments On Multiple M2-branes,''
  JHEP {\bf 0802} (2008) 105
  [arXiv:0712.3738 [hep-th]].


\bibitem{VanRaamsdonk:2008ft}
  M.~Van Raamsdonk,
  ``Comments on the Bagger-Lambert theory and multiple M2-branes,''
  arXiv:0803.3803 [hep-th].


\bibitem{Lambert:2008et}
  N.~Lambert and D.~Tong,
  ``Membranes on an Orbifold,''
  arXiv:0804.1114 [hep-th].

\bibitem{Distler:2008mk}
  J.~Distler, S.~Mukhi, C.~Papageorgakis and M.~Van Raamsdonk,
  ``M2-branes on M-folds,''
  arXiv:0804.1256 [hep-th].




\bibitem{Gomis:2008cv}
  J.~Gomis, A.~J.~Salim and F.~Passerini,
  ``Matrix Theory of Type IIB Plane Wave from Membranes,''
  arXiv:0804.2186 [hep-th].

\bibitem{Hosomichi:2008qk}
  K.~Hosomichi, K.~M.~Lee and S.~Lee,
  ``Mass-Deformed Bagger-Lambert Theory and its BPS Objects,''
  arXiv:0804.2519 [hep-th].

\bibitem{Bergshoeff:2008cz}
  E.~A.~Bergshoeff, M.~de Roo and O.~Hohm,
  ``Multiple M2-branes and the Embedding Tensor,''
  arXiv:0804.2201 [hep-th].

\bibitem{Berman:2008be}
  D.~S.~Berman, L.~C.~Tadrowski and D.~C.~Thompson,
  ``Aspects of Multiple Membranes,''
  arXiv:0803.3611 [hep-th].

\bibitem{Morozov:2008cb}
  A.~Morozov,
  ``On the Problem of Multiple M2 Branes,''
  arXiv:0804.0913 [hep-th].


\bibitem{Ho:2008nn}
  P.~M.~Ho and Y.~Matsuo,
  ``M5 from M2,''
  arXiv:0804.3629 [hep-th].

\bibitem{blau}
  D.~Birmingham, M.~Blau, M.~Rakowski and G.~Thompson,
  ``Topological field theory,''
  Phys.\ Rept.\  {\bf 209}, 129 (1991).

  \bibitem{verlinde}
  S.~Benvenuti, D.~Rodriguez-Gomez, E.~Tonni and H.~Verlinde,
  ``N=8 superconformal gauge theories and M2 branes,''
  arXiv:0805.1087 [hep-th].
  
  \bibitem{matsuo}
  P.~M.~Ho, Y.~Imamura and Y.~Matsuo,
  ``M2 to D2 revisited,''
  arXiv:0805.1202 [hep-th].
  
\end{thebibliography}
\end{document}